\documentclass[english,amssymb,floatfix]{revtex4}
\usepackage[T1]{fontenc}
\usepackage[latin9]{inputenc}
\usepackage{amsmath}
\usepackage{amssymb}
\usepackage{graphicx}
\usepackage{esint}

\makeatletter

\@ifundefined{textcolor}{}
{%
 \definecolor{BLACK}{gray}{0}
 \definecolor{WHITE}{gray}{1}
 \definecolor{RED}{rgb}{1,0,0}
 \definecolor{GREEN}{rgb}{0,1,0}
 \definecolor{BLUE}{rgb}{0,0,1}
 \definecolor{CYAN}{cmyk}{1,0,0,0}
 \definecolor{MAGENTA}{cmyk}{0,1,0,0}
 \definecolor{YELLOW}{cmyk}{0,0,1,0}
}

\usepackage{amsfonts}\usepackage{bbm}\usepackage{epsfig}\usepackage{graphics}%\textheight 24.0cm
\usepackage{ifpdf}\ifpdf%if using pdfLaTeX in PDF mode
\DeclareGraphicsExtensions{.pdf,.png}\usepackage{pgf}\usepackage{tikz}\usepackage{epstopdf}

\else%if using LaTeX or pdfLaTeX in DVI mode
\DeclareGraphicsExtensions{.eps,.png}
\fi

\graphicspath{{./}}

\newcommand{\lsim}{\,\lower2truept\hbox{${<\atop\hbox{\raise4truept\hbox{$\sim$}}}$}\,}\newcommand{\gsim}{\,\lower2truept\hbox{${>\atop\hbox{\raise4truept\hbox{$\sim$}}}$}\,}

\newcommand{\be}{\begin{equation}}\newcommand{\ee}{\end{equation}}\newcommand{\bea}{\begin{eqnarray}}\newcommand{\eea}{\end{eqnarray}}\newcommand{\beann}{\begin{eqnarray*}}\newcommand{\eeann}{\end{eqnarray*}}

\newcommand{\eprint}[1]{\url{arXiv:#1}}

\makeatother

\usepackage{babel}

\makeatother

\usepackage{babel}

\makeatother

\usepackage{babel}

\makeatother

\usepackage{babel}

\makeatother

\usepackage{babel}

\begin{document}

\title[How early is early dark energy?]
{How early is early dark energy?}

\author{Valeria Pettorino$^{1}$, Luca Amendola$^{2}$, Christof Wetterich$^{2}$}
\affiliation{
$^1$ D\'epartement de Physique Th\'eorique and Center for Astroparticle Physics, Universit\'e de Gen\`eve, 24 quai Ernest Ansermet, CH--1211 Gen\`eve 4, Switzerland.
\\
$^2$ Institut f\"ur Theoretische Physik, Universit\"at Heidelberg,
Philosophenweg 16, D-69120 Heidelberg, Germany.
}

\begin{abstract}
We investigate constraints on early dark energy (EDE) from the Cosmic Microwave
Background (CMB) anisotropy, taking into account data from WMAP9 combined with
 latest small
scale
measurements from the South Pole Telescope (SPT). For a constant EDE fraction we
propose a new parametrization with one less parameter but still enough to provide
similar results to the ones previously studied in literature. The main
emphasis of our analysis, however, compares a new set of
different EDE parametrizations that reveal how CMB constraints 
depend on the redshift epoch at which Dark Energy was non negligible.  We find
that bounds on EDE get 
substantially weaker if dark energy starts to be non-negligible later, with
early dark energy fraction $\Omega_e$
free to go up to about $5\%$ at 2 sigma if the onset of EDE happens at $z \lsim
100$ . Tight bounds around $1-2\%$ are obtained whenever 
dark energy is present at last scattering, even if its effects switch off
afterwards. We show that the CMB mainly constrains the presence of Dark Energy at the time
of its emission, while EDE-modifications of the subsequent growth of structure
are less important.
\end{abstract}

\date{\today}
\maketitle

\section{Introduction}
Models of dynamical dark energy or quintessence \cite{wetterich_1988,
ratra_peebles_1988} can be roughly divided into two classes: with or without
early dark energy (EDE). Models without early dark energy have a cosmology that
is indistinguishable from a cosmological constant ($\Lambda$CDM) for a redshift
larger than a few. Dark energy simply plays no role in early epochs of the
universe. Usually, this class of models shares the same fine tuning and "why now"
problems as $\Lambda$CDM. On the other hand, models with early dark energy are
characterized by a non-negligible amount of dark energy at early times, that distinguishes them
from $\Lambda$CDM: they can be related to a scaling or attractor solution where
the fraction of dark energy follows the fraction of the dominant matter or
radiation component \cite{wetterich_1988}.  As a consequence of the scaling
behavior, such models predict in early cosmology a non-vanishing dark energy
fraction $\Omega_e$. Since $\Omega_e$ must be substantially smaller than the
present dark energy fraction this class of models needs an ``exit mechanism''
 explaining why the scaling solution ends in the recent cosmological past, such that the
fraction in dark energy increases subsequently,  leading to the observed
accelerated expansion epoch. A large class of models of this type   has been
proposed. An example is growing neutrino quintessence \cite{Amendola_etal_2008,
3A}. Such models may be very close to $\Lambda$CDM in the present and recent
cosmological epoch, but have a nonzero $\Omega_e$ as a distinctive feature.    

The central quantity for EDE is $\Omega_e$,  which measures the amount of dark
energy present at early cosmological epochs. It is therefore natural to use this
quantity for a parametrization of the time history of dark energy. This was done
in refs.\cite{CWP}, where the name ``early dark energy'' was  proposed
and in subsequent works along this line \cite{doran_robbers_2006}. 
Besides $\Omega_{e}$ these parametrizations use, as usual, $\Omega_m$ (from
which $\Omega_{de} = 1- \Omega_m -\Omega_r$ is derived) plus the present dark energy
equation of state parameter $w_0$.
While the equation of state puts emphasis on how fast the dark energy fraction
$\Omega_{de}(z)$ gets small as $z$ increases, $\Omega_e$ measures how much  dark
energy is present at high $z$. 
The parametrization used in refs.\cite{doran_robbers_2006, calabrese_etal_2011,
reichardt_etal_2012, act_2013arXiv1301.0824S} (EDE1) is illustrated in the left panel of Fig.
\ref{fig:omega_ede}. This behavior differs from $\Lambda$CDM, that can be seen
as the approximate limit in which $\Omega_e \rightarrow 0$. Measuring how $\Omega_e$ differs
from zero represents therefore a valuable tool to distinguish, and possibly
falsify, a cosmological constant scenario from dynamical dark energy. 

Bounds on the value of $\Omega_e$ have been found from various observations as
nucleosynthesis \cite{wetterich_1988, BS, BHM}, structure formation \cite{FJ2,
FJ1, DSW} or the separation of peaks in the CMB anisotropies \cite{DLSW}. The
most precise bounds on EDE (or ``early quintessence'' as used in \cite{DSW, Ca})
arise from the analysis of CMB-anisotropies \cite{DLW, Ca, calabrese_etal_2011,
reichardt_etal_2012, act_2013arXiv1301.0824S}. In particular, very severe bounds $\Omega_e \lesssim 0.02$
at $95\%$ confidence for EDE1 have been found in the analysis of SPT-data by
Reichard et al. \cite{reichardt_etal_2012}, a factor of three improvement over
WMAP7 data alone. Similarly, the recent \cite{act_2013arXiv1301.0824S} limits $\Omega_e < 0.025$ from WMAP7+ACT+ACTDefl at 95$\%$ confidence level. Such an improvement mostly comes from including CMB data at
small angular scales, extending previous measurements of the temperature power
spectrum (WMAP7, \cite{Komatsu2011}) down to $\ell \sim 3000$, with first
compelling evidence of CMB lensing \cite{k11, act_2011}.
The impact of small-scale CMB measurements and gravitational lensing on
cosmology is quite significant \cite{lewis_challinor_2006} and can indeed be
used to constrain cosmological parameters. In particular, CMB lensing can be
used to distinguish among different dark energy models \cite{verde_spergel_2002,
giovi_etal_2003, acquaviva_baccigalupi_2006, 2004PhRvD..70b3515A, hu_etal_2006,
Sherwin:2011gv}. 

Using updated data from WMAP \cite{wmap9} and SPT \cite{spt_hou_etal_2012}, we perform here an analysis similar to ref.\cite{reichardt_etal_2012} with emphasis on which cosmological period or
redshift range the bounds apply: that is to say, how early is early? We confirm
the analysis of \cite{reichardt_etal_2012} if $\Omega_e$ is constant for all
redshifts  $z\gtrsim 3$, while  we find that the bounds are considerably weaker if
the effect of EDE is restricted to a limited range in $z$. After last
scattering the main effect of EDE is a reduced growth of structure. This could be compensated by an enhancement of
the growth rate due to other phenomena.   An example is the growth of neutrino
lumps in growing neutrino quintessence \cite{Wintergerst:2009fh,
Pettorino:2010bv}. Thus bounds on EDE for the period after last scattering have
to be handled with care. On the other hand, a presence of dark energy during the
period of last scattering mainly influences the evolution of geometry with much
less possibilities of compensation.

Early dark energy parametrizations grab  features of a large class of
dynamical dark energy models, namely having a non-negligible fraction of
dark energy at early times. The amount of early dark energy influences CMB peaks
in various ways and can be strongly constrained when including small scale
measurements. We here investigate whether CMB constraints are affected by a
variation of the epoch when $\Omega_e$ is non negligible. In particular, we aim 
at distinguishing between two EDE-effects. The first is the reduced structure
growth in the period after last scattering. This implies a smaller number of
clusters as compared to $\Lambda$CDM, and therefore a weaker lensing potential
influencing the anisotropies at high $\ell$. We ``isolate'' this effect by
``switching on'' EDE only after last scattering, at a scale factor $a_e$. This
is achieved by our parametrization EDE3. The second effect is the influence on
the position and height of peaks arising from dark energy present at the epoch
of last scattering \cite{DLSW}. This effect is ``isolated'' by our
parametrization EDE4 for which EDE is present until a  final scale factor $a_f$
and switched off afterwards. Of course, the ``isolation'' is not perfect - the
different parametrizations merely put particular emphasis on certain features. 

We discuss in detail the differences between four EDE parametrizations,
analyzing the effect of EDE on the CMB temperature power spectrum for each of
them. The first parametrization (EDE1) is the one adopted by
\cite{doran_robbers_2006}  and tested in \cite{reichardt_etal_2012}. The second,
new, parametrization (EDE2) uses only two parameters instead of three, namely
$\Omega_e$ and $\Omega_{m0}$,   without using $w_0$ as a third parameter as it
happens for EDE1. It keeps $\Omega_e$ constant in the past, as   for EDE1, but
has a sharper transition from a non negligible EDE to $\Lambda$CDM.
Incidentally, we note that such a sharp transition is effectively realized
 for
 growing neutrino quintessence cosmologies \cite{Amendola_etal_2008, Mota:2008nj,
Pettorino:2010bv}. We find only small differences between EDE1 and EDE2,  despite the absence of $w_0$ in EDE2. This
also shows that the influence of the equation of state $w_0$ on the bound for
$\Omega_e$ is small. The third and fourth parametrizations (EDE3 and EDE4), on
which most of our analysis and results are focused, are also proposed here for
the first time and they are used as a simple, yet powerful, tool to evaluate
which epoch is most relevant for EDE constraints. 

This paper is organized as follows. In Section II we review early dark energy
and introduce the parametrizations used throughout the paper. In Section III we
recall the basics of CMB lensing and the effects of EDE on the CMB lensing
potential. In Section IV we illustrate the methods used, MonteCarlo simulations,
parameters adopted and data. Our results are described in section V and
conclusions drawn in Section VI.

\section{Early Dark Energy}
\begin{figure*}
        \label{fig:omega_ede_EDE1}
   \includegraphics[width=8.cm]{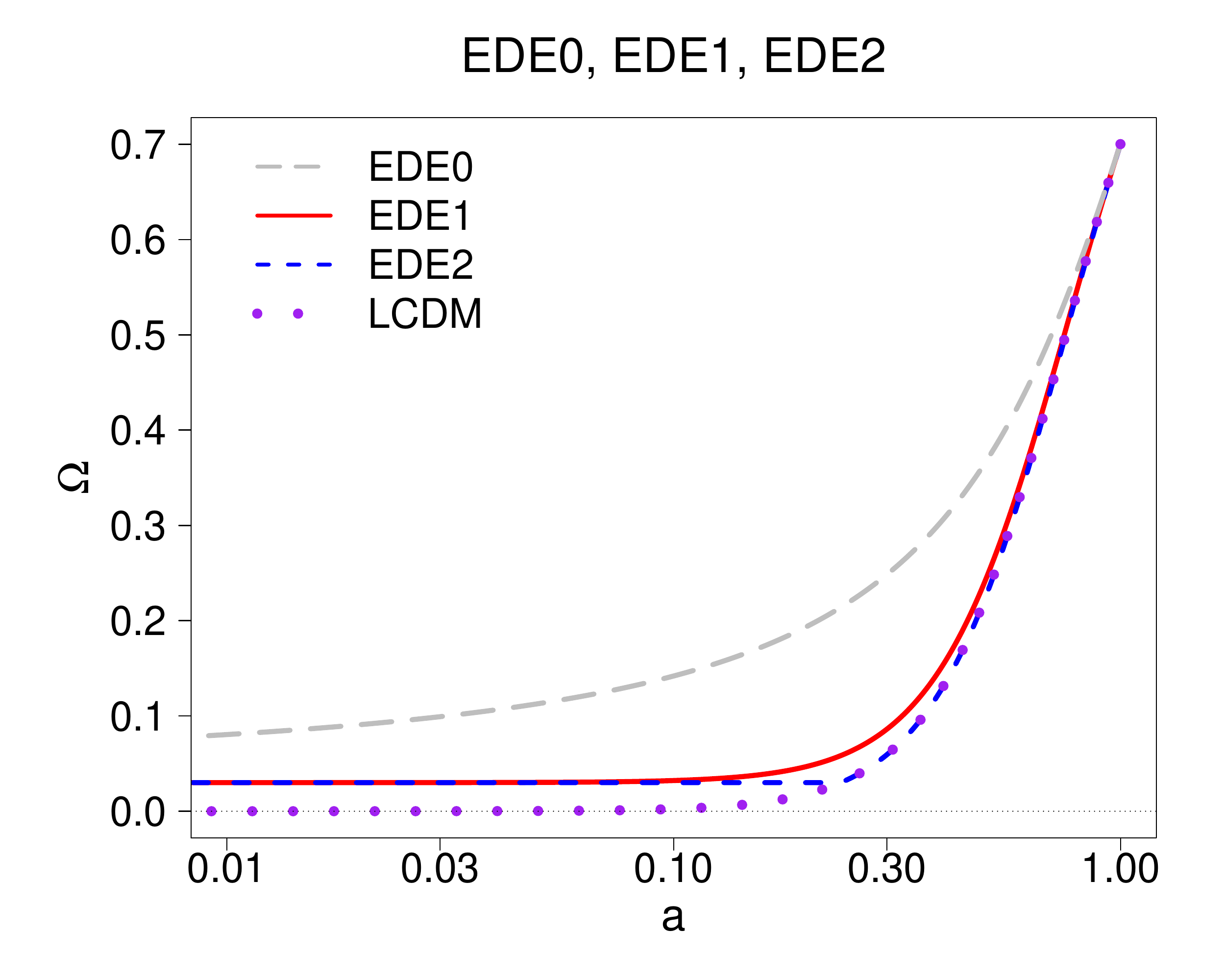}
   \includegraphics[width=8.cm]{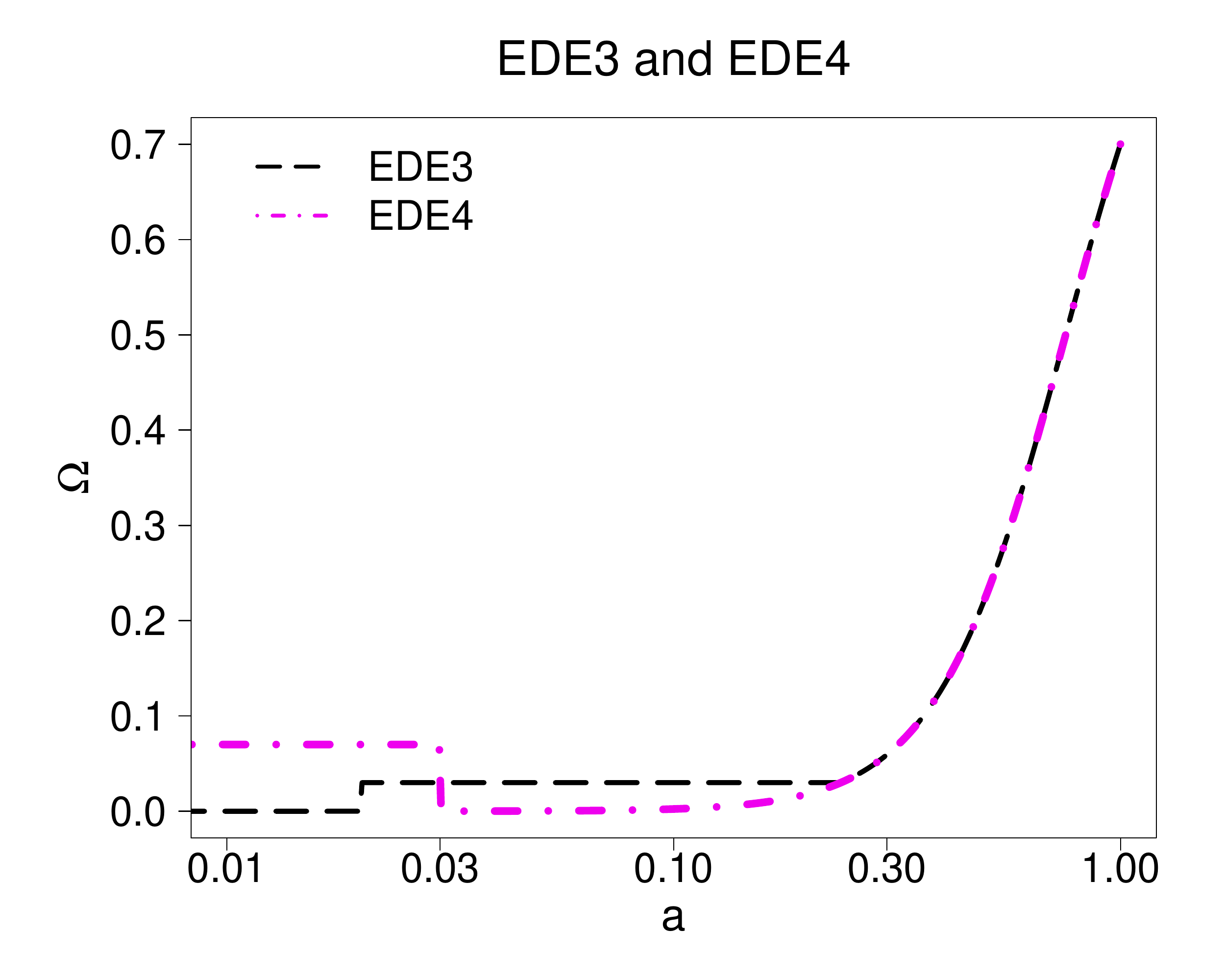}
     \caption{\small Dark energy density ratio with respect to the critical
density vs the scale factor $a$. Left panel: EDE0 (long dashed grey),  EDE1 (solid
red) and EDE2 (dashed blue) for the same choice $\Omega_{e}= 0.03$. $\Lambda$CDM
is also shown for comparison (dotted purple). Right panel: EDE3 (solid black),
with $\Omega_e = 0.03$ and $a_e = 0.02$ is shown together with EDE4 (dot-dashed
pink) for $a_f = 0.03 < a_c$ and $\Omega_e = 0.07$. }
             \label{fig:omega_ede}
\end{figure*}

The presence of a non negligible amount of dark energy at early times is a quite
generic feature present in several dynamical dark energy models. As such, it is
interesting to understand how much EDE can be present and at which epochs, given
our present knowledge of CMB measurements. Our aim is to evaluate the   impact
of EDE on CMB, identifying the effect that is has at different epochs of the
evolution of the Universe.  For this purpose, we consider here four classes of
different EDE models. 
In Fig.(\ref{fig:omega_ede}) we show the evolution of the dark energy fraction
for the four EDE parametrizations  plus EDE0, which corresponds to the first
proposal of an EDE parametrization \cite{CWP}. In Fig.(\ref{fig:w_ede}) we plot
the corresponding equation of state.

\normalsize

\begin{figure*}
   \includegraphics[width=8.cm]{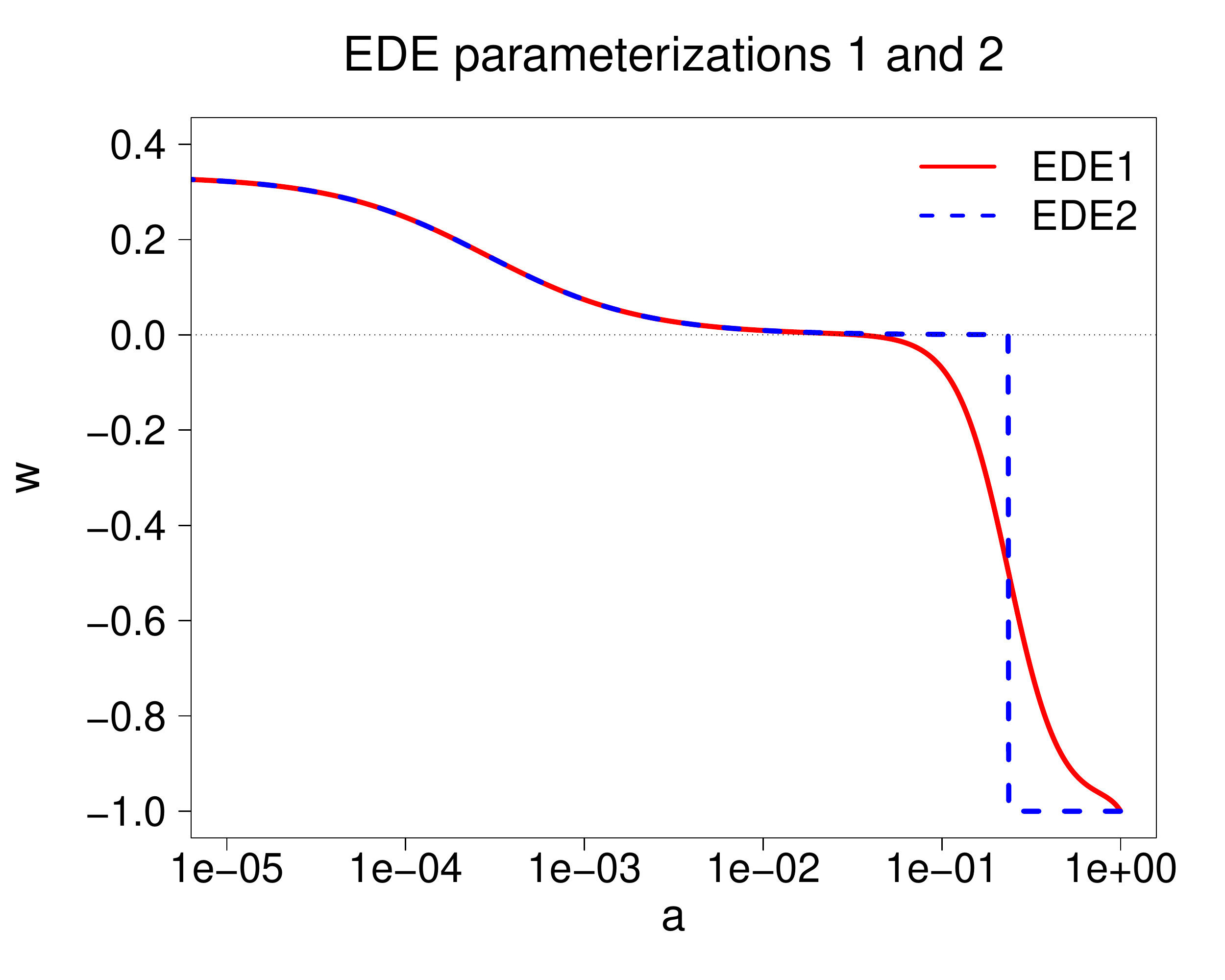}
   \includegraphics[width=8.cm]{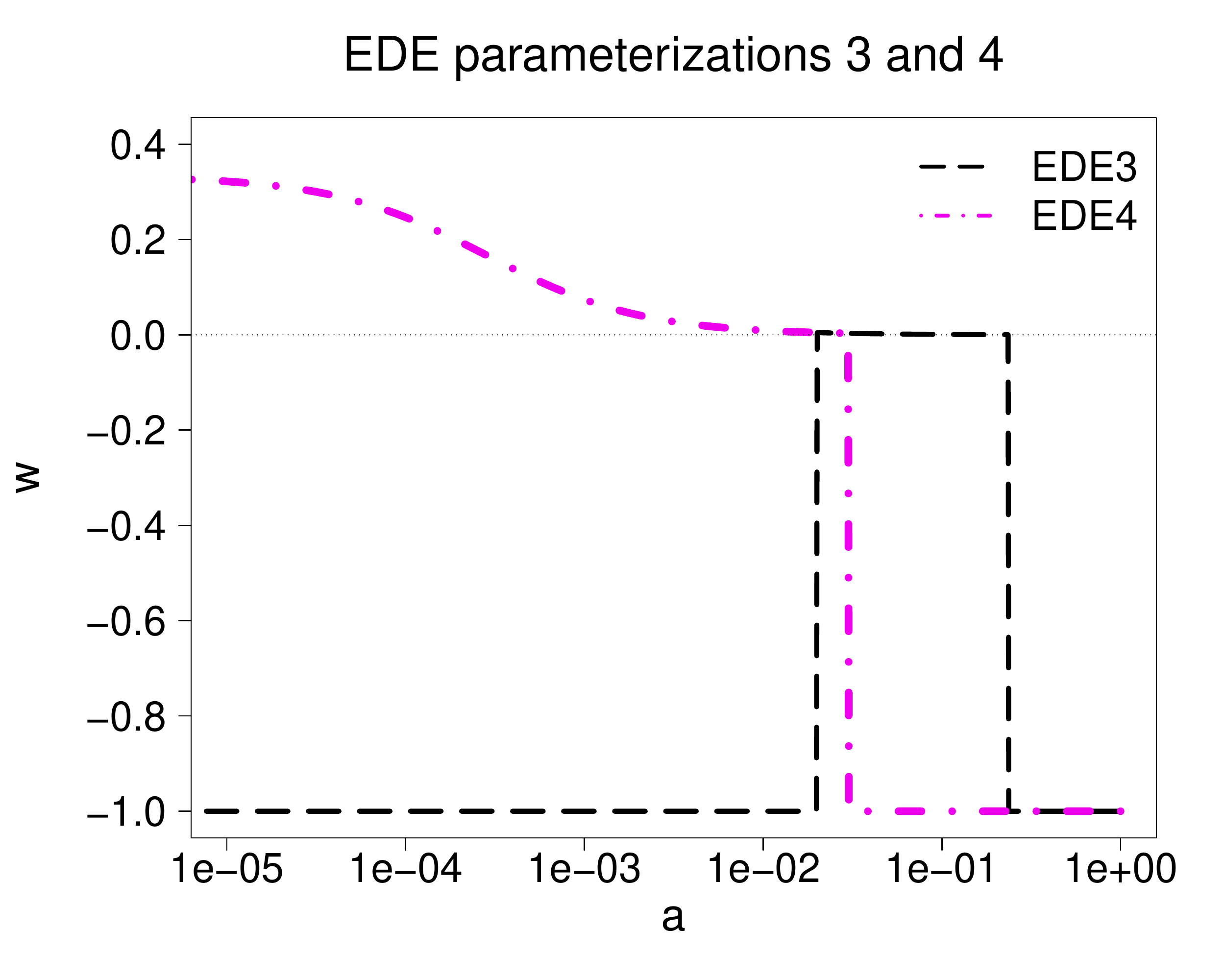}
\label{fig:w_ede}
     \caption{\small Dark energy equation of state $w(a)$ as a function of the
scale factor $a$. Left panel: EDE1 (solid red) and EDE2 (dashed blue) for the same
choice of the $\Omega_{e}$ parameter ($\Omega_e = 0.03$). Right panel: EDE3
(long dashed black) and EDE4 (dot-dashed magenta) for $\Omega_{e} = 0.03$ and 
$\Omega_{e} = 0.07$ respectively. }
\label{fig:w_ede}
\end{figure*}
\normalsize

\subsection{EDE 0}
This is the first EDE parametrization which was proposed in \cite{CWP}. Here 
\begin{eqnarray}\label{wde1-a}
\Omega_{de}(a) = \frac{e^{R(y)}}{1+e^{R(y)}}  \,\,\,   ,
\end{eqnarray}
where $y \equiv -\ln{a}$ and R(y) is obtained as 
\begin{eqnarray}\label{wde1-b}
R(y) = R_0 + \frac{3 w_0 y}{1 + b y} \,\,\,   .
\end{eqnarray}
The constant $R_0$ is directly related to $\Omega_M$ by:
\begin{eqnarray}\label{wde1-c}
R_0 = \ln{\left(\frac{1-\Omega_M}{\Omega_M}\right)} \,\,\,   ,
\end{eqnarray}
$w_0$ is the present equation of state for dark energy and $b$ is a parameter
that can be related to the amount of early dark energy $\Omega_e$:
\begin{eqnarray}\label{wde1-d}
b = -\frac{3 w_0}{\ln{\left(\frac{1-\Omega_e}{\Omega_e}\right)} +
\ln{\left(\frac{1-\Omega_M}{\Omega_M}\right)}} \,\,\, .
\end{eqnarray}
In the limit in which $\Omega_e \rightarrow 0$ and $a \rightarrow 1$, the
parameter $b \rightarrow 0$ and the model reduces to $\Lambda$CDM. If $b \neq 0$
there is a very smooth transition
from $\Lambda$CDM to a model with a non negligible amount of dark energy at
early times. The behavior of $\Omega_{de}$ for this parametrization is plotted
in Fig.\ref{fig:omega_ede}. We will not use this parametrization in the
following but it is interesting to compare its behavior with the sharper
transition that characterizes the other parametrizations described below.

\subsection{EDE 1}
The first parametrization  that we investigate numerically (EDE1) is  the one proposed in 
\cite{doran_robbers_2006} and tested in refs.\cite{calabrese_etal_2011, reichardt_etal_2012, act_2013arXiv1301.0824S}.
The dependence of the dark energy fraction, $\Omega_{de}$, on the scale
parameter $a$ is given by
\begin{eqnarray}\label{wde1}
\Omega_{de}(a) = \frac{\Omega_{de}^{(0)}-\Omega_e (1-a^{-3
w_0})}{\Omega_{de}^{0}+\Omega_m^0 a^{3 w_0}}+\Omega_e(1-a^{-3 w_0}). 
\end{eqnarray}
Eq. \eqref{wde1} uses three parameters, the present matter fraction
$\Omega_{m0}$, the early dark energy fraction $\Omega_e$ and the present dark
energy equation of state $w_0$, with $\Omega_{de0}=1-\Omega_{m0}$. For any given
function $\Omega_{de}(a)$ the scale dependent equation of state $w(a)$ obtains
as a simple derivative \cite{CWDW}
\begin{eqnarray}\label{wde}
w(a)=-\frac{1}{3[1-\Omega_{de}(a)]}\frac{d \ln{\Omega_{de}}}{d
\ln{a}}+\frac{a_{eq}}{3(a+a_{eq})},
\end{eqnarray}
where $a_{eq} \sim 3200$ is the scale factor at matter-radiation equality, while
 the energy density is given by
\begin{eqnarray}\label{rhode}
\rho_{de}(a)=\frac{\rho_{de0}}{a^{3}} \frac{\Omega_{de}}{\Omega_{de0}}\frac{\Omega_{de}-1}{\Omega_{de0}-1}\left(1+\frac {a_{eq}}{a}\right)\frac{1}{1+a_{eq}}
\,.
\end{eqnarray}

\subsection{EDE 2}
The second parametrization (EDE2), that we propose here for the first time,
reads as follows:
\begin{eqnarray}\label{3}
\Omega_{de}(a) = \left\{
\begin{array}{c l}      
    \Omega_{e} & a < a_c  \,\,\, ,\\
    \frac{\Omega_{de0}}{\Omega_{de0}+\Omega_{m0} a^{-3}+\Omega_{r0} a^{-4}} & a
\ge a_c      \,\,\, .
\end{array}\right.
\end{eqnarray}
Here $a_c$ is determined by continuity at $a_c$, such that (neglecting $\Omega_{r0}$):
\begin{eqnarray}
a_c = \left[ \frac{\Omega_e \Omega_{m_0}}{\Omega_{de0}
(1-\Omega_e)}\right]^{1/3}  \,\,\, .
\end{eqnarray}
In this way $\Omega_e$ is the only additional parameter, beyond $\Omega_{m0}$.
This is a minimal parametrization of EDE. Similar to other two-parameter
settings for dark energy, as the ones using $\Omega_{m0}$ and $w_0$, it is
useful if data allow only for a rough distinction of dynamical dark energy from
$\Lambda$CDM. This second parametrization considers a somewhat sharper
transition between a phase in which there is a constant $\Omega_{e}$
contribution and the epoch in which dark energy looks close to a cosmological
constant.  We recall that in models of growing neutrino quintessence \cite{Amendola_etal_2008,3A,
Mota:2008nj, Pettorino:2010bv} explain the ``why now'' problem by a cosmological
trigger event, namely neutrinos becoming non-relativistic; for such models one
typically finds a rather sharp transition between the two epochs. 

\subsection{EDE 3}
For the EDE3-parametrization EDE becomes important only for $a>a_e$. Beyond $\Omega_{m0}$ it has two parameters, $\Omega_e$ and $a_e$, according to 
\begin{eqnarray}
\Omega_{de}(a) = \left\{
\begin{array}{c ll} 
     \frac{\Omega_{de0}}{\Omega_{de0}+\Omega_{m0} a^{-3}+\Omega_{r0} a^{-4}}   &   &  a \le a_e\\     
\\
    \Omega_{e} & &  a_e < a < a_c\\
\\
    \frac{\Omega_{de0}}{\Omega_{de0}+\Omega_{m0} a^{-3}+\Omega_{r0} a^{-4}} & &  a>a_c \,\,\, .
\end{array}\right.
\end{eqnarray}
In this case, early dark energy is present in the time interval between $a_e < a
< a_c$ while outside this interval it behaves as in $\Lambda$CDM.
In that interval, there is a non negligible EDE contribution, whose amount is
parametrized by $\Omega_e$.
As in EDE2, $a_c$ is fixed by the continuity condition, so that the parameters
characterizing this case are ($\Omega_e, a_e$), that is to say how much EDE
there is and how long its presence lasted. We note that for $a_e\ll a_c$
one has $\Omega_{de}(a\leq a_e)\approx 0$.

\subsection{EDE 4}
The next parametrization we choose provides somehow complementary information with respect to EDE3. 
For the EDE4 parametrization,

\begin{eqnarray}\label{6}
\Omega_{de}(a) = \left\{
\begin{array}{c l} 
    \Omega_{e}    & a < a_f\\     
    \frac{\Omega_{de0}}{\Omega_{de0}+\Omega_{m0} a^{-3}+\Omega_{r0} a^{-4}} & a>a_f  \,\,\, ,
\end{array}\right.
\end{eqnarray}
 early dark energy is present from early times until the scale factor reaches
the value $a_f$, while $\Lambda$CDM is recovered at later times. For $a_f\ll
a_c$ the dark energy fraction almost drops to zero at $a_f$, as visible in Fig.
\ref{fig:omega_ede}. To be formally consistent, we use eq. \eqref{6} only for
$a_f<a_c$, while for $a_f>a_c$ we employ EDE2, eq. \eqref{3}. We are interested,
however, only in the case $a_f<a_c$. The parameters characterizing this case are
($\Omega_e, a_f$), that is to say how much EDE there is and since when it
started to be negligible.

\section{CMB temperature power spectrum and lensing potential}

\begin{figure*}
   \includegraphics[width=15 cm]{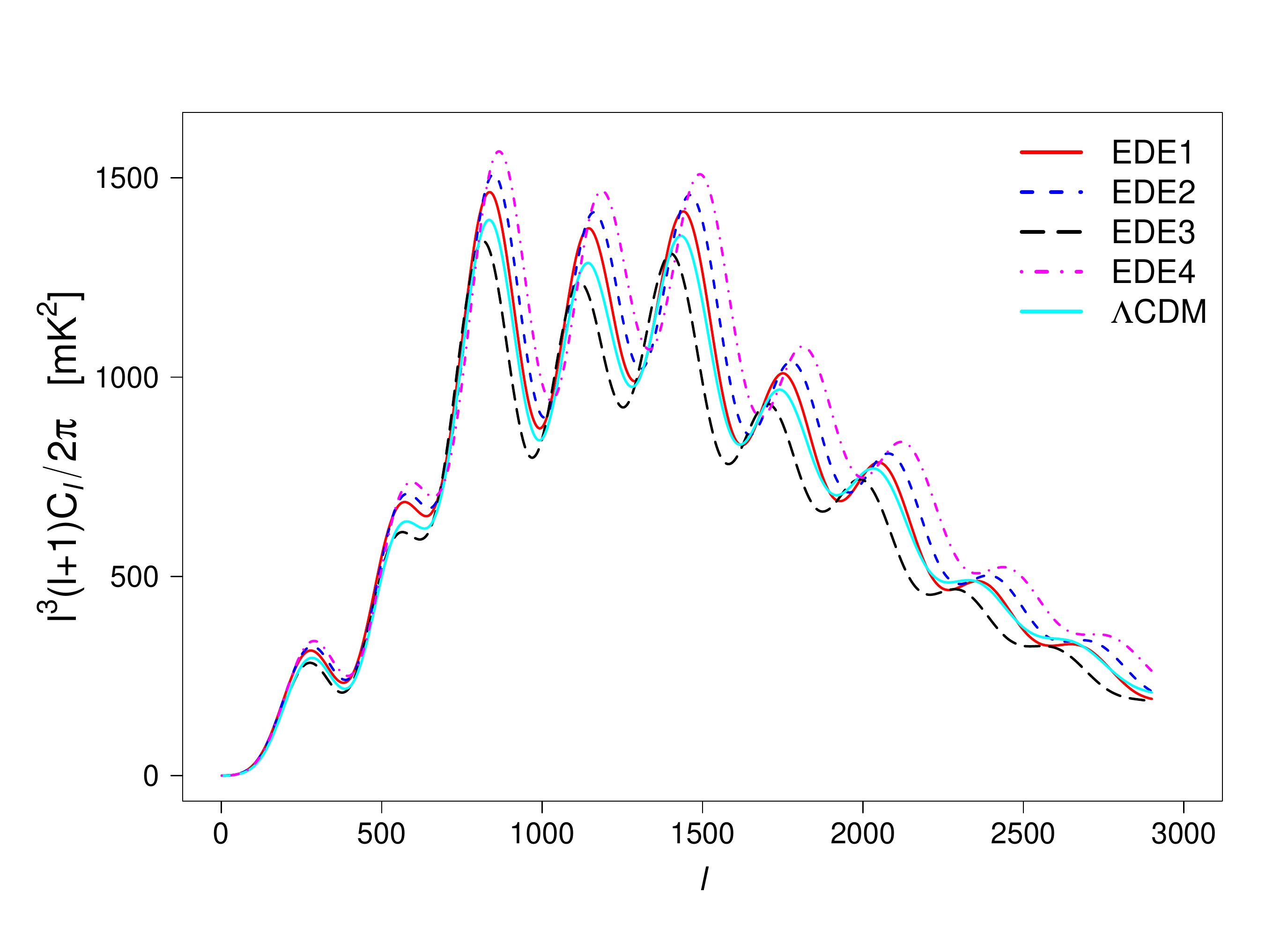}%
    \caption{\small Temperature CMB power spectrum for four different EDE
parametrizations. Here, in all cases, we have used a large value $\Omega_e=0.1$
to show the effect of the different parametrizations. For smaller values, as
$\Omega_e = 0.1$, the effect on the $C_\ell$ is quite small and smaller than the
one on the lensing potential that we will see in Fig.\ref{fig:lensingpot_ede}. EDE3 and
EDE4 have $a_e = 0.02$ and $a_f = 0.03$ respectively.
The $\Lambda$CDM spectrum is also shown for reference, for the same value of
$H_0$ and $\Omega_{m0}$ today. Note that we are plotting $\ell^3 (\ell+1)C_\ell / 2\pi$
rather than $\ell (\ell+1)C_\ell / 2\pi$.}
\label{fig:cl_1dn1}
\end{figure*}

  An increase in EDE decreases the amount of CDM in a way that differs from what
one would obtain by simply increasing the number of radiation species, since
dark energy and radiation decay differently in redshift. There are two main
effects of EDE on the CMB spectrum. The first consists in the 
 influence on CMB peaks of the amount of dark
energy present at the epoch of last scattering, the second is the
suppression of
structure growth, implying a smaller number of clusters as compared to
$\Lambda$CDM. These two effects are
approximately isolated in the two parametrizations EDE3 and EDE4.
The effect of different EDE parametrizations on the CMB temperature power
spectrum is shown in Fig.(\ref{fig:cl_1dn1}). For good visibility we have chosen
a large value $\Omega_e=0.1$. For more realistic values as $\Omega_e=0.03$ the
difference to $\Lambda$CDM is barely visible. The second parametrization, EDE2,
boosts the amplitude of peaks somewhat higher than for EDE1.

EDE in general also moves the peaks to higher or smaller multipoles
since they change the Hubble function $H(z)$ (see \cite{DLSW} for a
quantitative analysis). EDE1, EDE2 and EDE4 produce a
$H(z)$ always higher or equal to $\Lambda$CDM, both before and after
decoupling. This reduces both the sound horizon at decoupling and the
angular-diameter distance to last scattering, but the angular-diameter
distance reduction is proportionally smaller because the last phase is
in all cases identical to $\Lambda$CDM. The net effect on the peak
angular scale (which is essentially the ratio of the sound horizon to
the CMB distance) is therefore to move it to smaller values, or higher
multipoles. In the case of EDE3 with $a_e$ larger than the decoupling
epoch, however, the sound horizon does not change at all with respect
to $\Lambda$CDM and therefore the only effect is a decrease of the
angular-diameter distance to last scattering, which leads to smaller
peak multipoles. 
This can be unserstood also in terms of the quantitative formula (3) in ref. \cite{DLSW}: the presence of
EDE at the time of last scattering ($\Omega_{lss}=\Omega_e$ for EDE1,2,4) shifts the peaks to
higher $\ell$s, while the presence of EDE in the period afterwards increases the averaged equation of state $\bar{w}_0$ and shifts the peaks to lower $\ell$s.
This effect suggests that bounds on EDE will not
only depend on some time averaged fraction of dark energy, but also on the
detailed time history.

We discuss
here briefly another potential effect, namely the reduction of the growth rate
due to the presence of EDE. This results in a modified lensing potential.

\normalsize

%\section{CMB-lensing}
The CMB coming from the last scattering surface (LSS) is bent by gravitational
structures on the path towards us; this effect is called CMB-lensing 
\citep{Bartelmann:1999yn, lewis_challinor_2006}  and its observation has
recently been claimed by the ACT and SPT teams \cite{act_2011, k11, spt_2012}.
The deflection angle is of the order of 2 arcminutes, which would correspond to
small scales and $l > 3000$ multipoles, where CMB peaks are already damped by
photon diffusion.  However, deflection angles are correlated over degree scales,
so that lensing can modify the scales of the acoustic peaks, the main effect
being a smoothing of primary peaks and a transfer of power to larger multipoles.
CMB gravitational lensing naturally depends on the growth of perturbations and
on the gravitational potentials. As a consequence, it  is a good way to probe
the existence of dark energy \citep{verde_spergel_2002, giovi_etal_2003,
2004PhRvD..70b3515A, acquaviva_baccigalupi_2006, hu_etal_2006, Sherwin:2011gv,
Pettorino:2012ts, amendola_etal_2012} and has recently been used to reject
$\Omega_{\Lambda CDM}$ = 0 at more than $5 \sigma$ from CMB alone
\cite{spt_2012}. Since EDE has a strong impact on the growth of structure CMB
lensing  can be a sensitive probe to it \cite{reichardt_etal_2012}.

  We recall here only the main features of CMB-lensing, which is not the main
topic of this paper but quite important when dealing with EDE constraints. For
more details we refer to \cite{lewis_challinor_2006}. In the Matter Dominated
Era (MDE) the potentials encountered along the way are constant in the linear
regime and the gradient of the potential causes a total deflection angle
$\alpha$ given by:
\begin{equation}
\alpha = -2 \int_0^{\chi_{*}} d \chi \frac{f_K(\chi_* - \chi)}{f_K(\chi_*)} \nabla_{\perp} \Psi (\chi \hat{n}; \tau_0 - \chi),
\label{alpha_lensing}
\end{equation} 
where $\chi_*$ is the conformal distance of the source acting as a lens, $\Psi$ is its gravitational potential, $\eta_0 - \chi$ is the conformal time at which the CMB photon was at position $\chi \hat{n}$. Here the function $f_K(\chi)$ is the angular diameter distance and it's equal to $\chi$ for a flat Universe, while it's a function of the curvature parameter $K$ for non flat cosmologies \cite{lewis_challinor_2006}.

\begin{figure*}
\centering
   \includegraphics[width=14.cm]{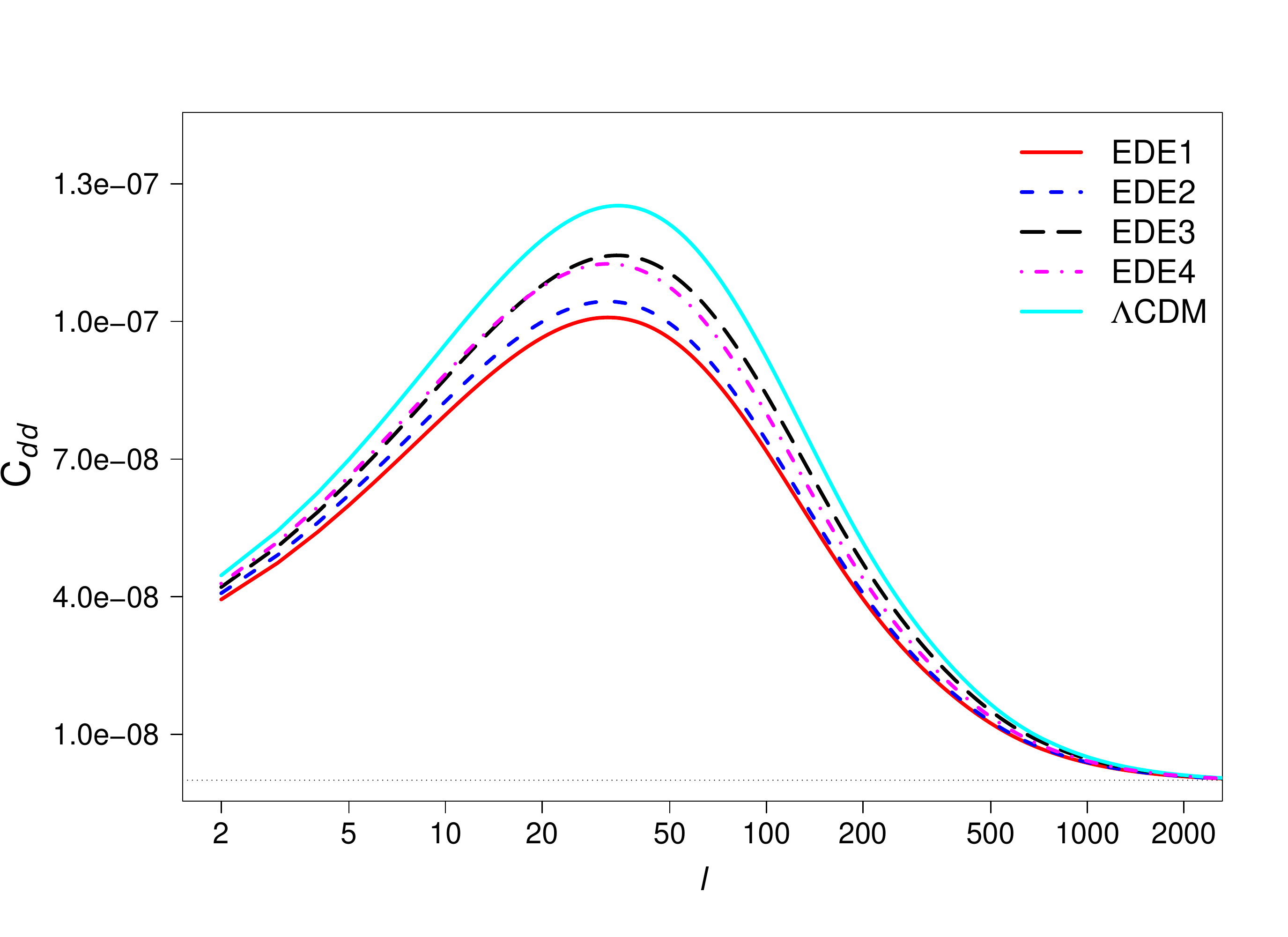}%
     \caption{\small Dimensionless lensing potential $C_{dd} \equiv
[\ell(\ell+1)]^2 C_\ell^{\psi}/(2\pi)$ versus multipole $\ell$ for different
parametrizations. All EDE parametrizations correspond to a value of $\Omega_e =
0.03$, which is enough to see the differences in the lensing potential. EDE3 and
EDE4 have $a_e = 0.02$ and $a_f = 0.03$ respectively. $\Lambda$CDM is also plotted for
comparison (top-solid cyan).}
\label{fig:lensingpot_ede}
\end{figure*}

The gravitational lensing  potential $\psi_\ell$ is defined by
\begin{equation}
\psi_\ell(\hat{n}) \equiv -2 \int_0^{\chi_*} d \chi
\frac{f_K(\chi_*-\chi)}{f_K(\chi_*)f_K(\chi)} \Psi(\chi \hat{n}; \eta_0 - \chi) 
\,\,\, .
\end{equation}
The lensed CMB temperature $\tilde{T}_{\hat{n}}$ in a direction $\hat{n}$ is
given by the unlensed temperature in the deflected direction $\tilde{T}(\hat{n})
= T(\hat{n}') = T(\hat{n} + \alpha)$ where at lowest order the deflection angle
$\alpha = \nabla \psi_\ell$ is just the gradient of the lensing potential.
Expanding the lensing potential into spherical harmonics, one can define also
the angular power spectrum $C_\ell^\psi$ corresponding to the lensing potential,
defined as $<\psi_{\ell m} {\psi^*}_{\ell'm'} > = \delta_{\ell \ell'}
\delta_{mm'} C_\ell^{\psi}$.  In Fig. \ref{fig:lensingpot_ede} we show $C_{dd} \equiv
[\ell(\ell+1)]^2 C_\ell^{\psi}/(2\pi)$ for different EDE parametrizations. 
Early Dark Energy reduces the lensing potential since structures  grow less
rapidly. For a given value $\Omega_e$ the reduction is stronger if EDE is present for a longer time, e.g. EDE1 and EDE2.

As we
can see from Fig. \ref{fig:lensingpot_ede}, the CMB lensing potential mainly gives contribution to
large scales or small $\ell$ up to $\ell \sim 1000$ or less.   However, the
lensed CMB temperature power spectrum  depends on the convolution between the
lensing potential and the unlensed temperature spectrum (see
\cite{lewis_challinor_2006} for more details):
\begin{equation}
\tilde{C_\ell^{\Theta}} \approx (1-\ell^2 R^\psi) C_\ell^{\Theta} +
\int{\frac{d^2 {\bf \ell}'}{(2 \pi)^2} [{\bf \ell}' ({\bf \ell}-{\bf \ell}')]^2
C^\psi_{|{\bf \ell}-{\bf \ell}'|} C_{\ell'}^\Theta} \,\,\,\, ,
\end{equation}
where $R^\psi = 1/2  <|\nabla \psi|^2>$ is the total deflection angle power
(typically of the order of $\sim 10^{-7}$), $\ell$ is the multipole,
$\tilde{C}_\ell^{\Theta}$ is the lensed temperature power spectrum,
${C}_\ell^{\Theta}$
is the unlensed temperature power spectrum and ${C}_\ell^{\psi}$ is the lensing
potential. The resulting effect is of several percent at $\ell > 1000$, thus
being important when estimating the spectrum up to small scales of $\ell \sim
3000$  as it happens for the SPT considered here. Also, the lensing depends on
time, combining information from decoupling (from the last scattering surface of
the CMB) and $z < 5$ (when large scale structures formed).

\normalsize

\section{Methods} \label{sec:methods}
We have implemented each EDE parametrization in CAMB, joined to COSMOMC, in
order to perform a Monte Carlo analysis and get information on the EDE
parameters from present CMB measurements.
We have compared theoretical predictions of CMB spectra with two datasets. The
first includes WMAP9 temperature spectra \cite{wmap9}, which  updates results
included in \cite{reichardt_etal_2012} (we have first also checked that we
reproduced results in \cite{reichardt_etal_2012} using WMAP7). The second
includes the  latest South Pole Telescope spectrum \cite{spt_hou_etal_2012, spt_2012}. 
In order to use SPT data and compare our EDE results with the ones of
\cite{reichardt_etal_2012, spt_hou_etal_2012}, we have installed the likelihood provided by the SPT
team  \cite{spt_hou_etal_2012, k11} on the SPT website
\footnote{http://pole.uchicago.edu/public/data/keisler11/index.html} and
integrated it with the recommended version of cosmic
\cite{cosmomc_lewis_bridle_2002} (August 2011).  Bandpowers and foreground templates have been updated to \cite{spt_hou_etal_2012}. Whenever including SPT data we
also marginalize over the three nuisance parameters described in \cite{spt_hou_etal_2012, k11}: two
of them refer to Poisson point sources and clustered point sources; the third
one adds power from the thermal and kinetic Sunyaev-Zel'dovich (SZ) effects. All
these effects are relevant when small scales ($l \gsim 2000$) are included and
must be therefore taken into account when SPT data are used.

For EDE1 and EDE2, the baseline set of parameters includes $\Theta = {\Omega_b
h^2, \Omega_{c} h^2, \theta_s, \log{\cal{A}}, n_s, \tau}$ while $\Omega_{de0}$ is a
derived parameter. In addition, when EDE is included two more parameters are
added for EDE1: $w_0, \Omega_{e}$, while EDE2 has one additional parameter
$\Omega_e$. For EDE1, we also investigated potential degeneracies with dark
energy parameters: {\it case1} includes EDE1 with both WMAP and SPT; {\it case2} includes
WMAP only; in {\it case3} we do not include lensing; in {\it case4} we marginalize over
the effective number of relativistic species $N_{eff}$.

For EDE3 (EDE4), we fixed the optical depth to $\tau = 0.088$ in order to speed
up the runs and we did several runs for different (fixed) values of $a_e$
($a_f$).  This is enough for our purpose: we want to compare the two parametrizations and estimate the
effect of this extra parameter, which gives information on how EDE bounds depend on the specific time at which dark energy is present.
Data and parameters of different runs are summarized in Tab.\ref{tab:runs}. 

 \begin{center}
\begin{table}
\begin{tabular}{cccccccc}  
Run & \begin{minipage}{100pt}
\center
CMB-Lensing \\ 
\end{minipage} &  
\begin{minipage}{50pt}
\center
WMAP9
\end{minipage} &
\begin{minipage}{50pt}
\center
SPT
\end{minipage} & 
\begin{minipage}{50pt}
\center
Parameters
\end{minipage} &
\\
\hline
 EDE1, case1   & \checkmark & \checkmark & \checkmark & baseline + $w_0$ + $\Omega_{e}$ \\
 EDE1, case2   & \checkmark & \checkmark & X & baseline + $w_0$ + $\Omega_{e}$\\
 EDE1, case3   & X & \checkmark & \checkmark & baseline + $w_0$ + $\Omega_{e}$ \\
 EDE1, case4   & \checkmark & \checkmark & \checkmark & baseline + $w_0$ + $\Omega_{e}$ + $N_{eff}$\\
  \hline
 EDE2   & \checkmark & \checkmark & \checkmark & baseline + $\Omega_{e}$ \\
 EDE3   & \checkmark & \checkmark & \checkmark & baseline + $\Omega_{e}$; fixed $\tau$ and $a_e$ \\
 EDE4   & \checkmark & \checkmark & \checkmark & baseline + $\Omega_{e}$; fixed $\tau$ and $a_f$ \\
\hline
\end{tabular} 
\caption{Data and parameters for runs investigated in this paper.  Note that unlike EDE1, the parametrizations EDE2, EDE3, EDE4 do not require the equation of state $w_0$ as an additional parameter.}  \label{tab:runs}
\end{table}  
\par\end{center}
\normalsize

\section{Results} \label{sec:results}

\subsection{Parametrizations 1 and 2}

\begin{figure*}
\centering
  
\includegraphics[width=16.cm]{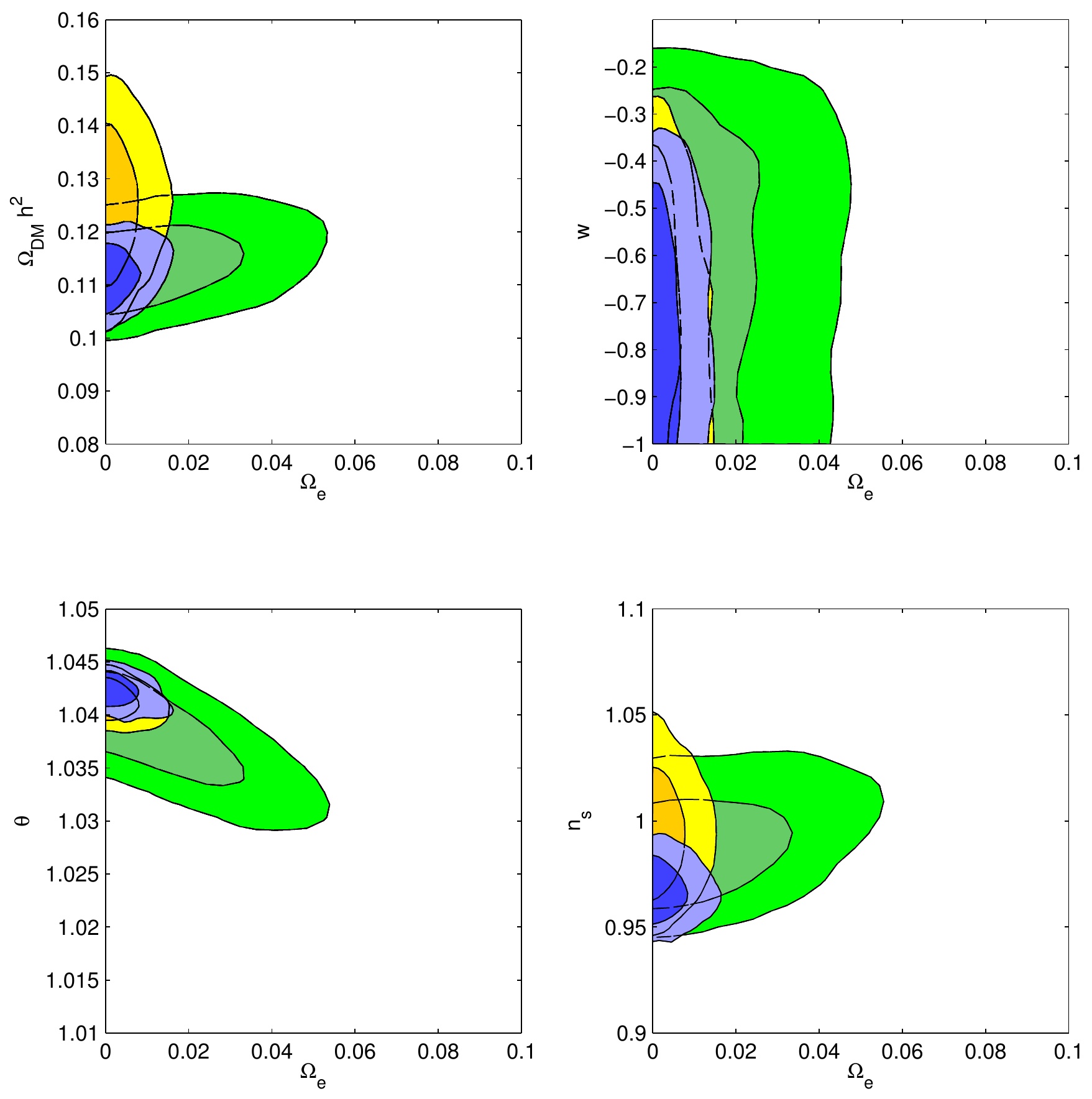}%
     \caption{\small Confidence contours for the cosmological parameters for
EDE1 models. We compare here run {\it case1} (WMAP9 + SPT, blue contours), {\it case2}
(WMAP9 only, green contours), {\it case4} (WMAP9 + SPT, $N_{eff}$, yellow contours).
$1 \sigma$ and $2 \sigma$ contours are shown.}
\label{fig:like2D_ede1ede2ede4}
\end{figure*}

In tables \ref{tab:ede1} and  \ref{tab:ede2} we compare the 1-sigma standard
deviation for baseline and derived parameters for the various runs listed in
Tab.\ref{tab:runs} for the first two EDE parametrizations.
The inclusion of SPT data ({\it case1}) improves the constraints on $\Omega_{e}$ with
respect to  WMAP9 only ({\it case2}), in agreement with results from
\cite{reichardt_etal_2012}  and updated to more recent data. This holds also for the second parametrization we
propose here, with no significant difference for most of the parameters. If the
second parametrization is adopted, however, constraints on derived parameters
like $\Omega_{de}$ (and  $\sigma_8$) improve of a factor of two, allowing though
for a weaker constraint on the age of the Universe.
No significant difference appears between {\it case1} and {\it case3} (no lensing) for the
first parametrization (Tab.\ref{tab:ede1}). 
When we allow for the number of relativistic species to vary from the reference
value $N_{eff} = 3.046$ we obtain results for the {\it case4} run: due to the
degeneracy between $N_{eff}$, the spectral index, dark energy and matter
parameters, constraints on $n_s$ are wider by almost a factor 2, constraints on
$\Omega_{c}h^2$ by a factor 2.5. Also, constraints on the $Y_{He}$ widen by a
factor of 41. 
No significant change, though, is seen on $\Omega_{e}$. Similar results hold for
both EDE parametrizations.
This is clearly visible also from the 2D likelihood plots in
fig.\ref{fig:like2D_ede1ede2ede4}  (also in agreement with \cite{reichardt_etal_2012} but updated to more recent data)
where we have added marginalization over $N_{eff}$. The 2D likelihood plot for
EDE2 looks very similar to the one for EDE1, so we do not show it.
In conclusion we note that   a steeper switch from
$\Lambda$CDM to non negligible dark energy, as in EDE2, does not affect present bounds on
$\Omega_e$. In the next paragraph we analyze results from EDE3 and EDE4, which
will help us understand which range in redshift determines the bounds on
$\Omega_e$.

 \begin{center}
\begin{table} 
\begin{tabular}{lllll} 
\textbf{Parameters and errors for EDE1} \\
\hline
\begin{minipage}{100pt}
\flushleft
Parameter  
\end{minipage} & 
\begin{minipage}{70pt}
\flushleft
case1 
\end{minipage} &  
\begin{minipage}{70pt}
\flushleft
case2 
\end{minipage} &
\begin{minipage}{70pt}
\flushleft
case3 
\end{minipage} &
\begin{minipage}{70pt}
\flushleft
case4 
\end{minipage} 
\\
\hline
$  \Omega_b h^2  $ &$0.022 \pm 0.00035$  & 		$0.023 \pm 0.00060 $ 	& $0.021 \pm 0.00033$ & $0.023 \pm 0.00049$ 	\\
$ \Omega_{c} h^2  $ & $ 0.11 \pm  0.0039$ & 		$0.11 \pm 0.0053$ 	 & $0.11 \pm 0.0041$	& $0.13 \pm 0.0087$ 		\\     
$ \theta_s$ & $1.04 \pm 0.0011$ 		  & 		$1.04 \pm 0.0033$ 	& $1.04 \pm 0.00094	$ & $1.04 \pm 0.0012 $	\\
   $ n_s$ & $0.967 \pm 0.0093 $ 			& 		$0.990 \pm 0.018$ 	& $0.950 \pm 0.0095$	& $0.995 \pm 0.018 $		\\
   $w_0$ &  $ < -0.44 $ 					& 		$< - 0.26 $ 		& $< -0.52 $	& $< -0.39$			\\  
   $\Omega_{e}$ &  $ < 0.013$ 		& 		$< 0.043 $		& $<  0.0098$	&  $< 0.012 $			\\  
   $ N_{eff}$ & -  					& - 					& -		& $3.9 \pm 0.51$ 			 \\
   $ Y_{He}$ & $0.25 \pm 0.00015 $ 		& 		$0.25 \pm 0.00025$ 	   & $0.25 \pm 0.00015$	& $0.26 \pm 0.0063 $	\\
   $ \Omega_{de0}$ & $0.64 \pm 0.069$ 	& 		$0.57 \pm 0.091$ 	& $0.65 \pm 0.056$	& $0.64 \pm 0.071$		\\
   $ H_0$ & $62.0 \pm 5.5$ 			& 		$57.7 \pm 6.2 $ 	& $62.9 \pm 4.6$	& $64.7 \pm 6.2$ 		 \\ 
\hline
\end{tabular}
\caption{ Mean value plus standard deviation for a selection of
baseline and derived parameters in EDE1. In the case of $\Omega_e$ and $w$ we
report the $95\%$ confidence upper limit. Values of $case1$ are compatible with
\cite{reichardt_etal_2012}. Limits on $\Omega_e$ do not seem to be sensitive to
marginalizing over the effective relativistic degrees of freedom $N_{eff}$. }
\label{tab:ede1}
\end{table}
\par\end{center} 
\normalsize

 \begin{center}
\begin{table} 
\begin{tabular}{lllll}
\textbf{Parameters and errors for EDE2} \\
\hline
\begin{minipage}{100pt}
\flushleft
Parameter  
\end{minipage} & 
\begin{minipage}{70pt}
\flushleft
case 1 
\end{minipage} &  
\begin{minipage}{70pt}
\flushleft
case 2 
\end{minipage} &
\begin{minipage}{70pt}
\flushleft
case 4 
\end{minipage} &
\\
\hline
$  \Omega_b h^2  $ &$0.022 \pm 0.00035$  & 		$0.023 \pm 0.00051 $	& $0.023 \pm 0.00047$ 	\\
$ \Omega_{c} h^2  $ & $ 0.11 \pm  0.0039$ & 		$0.12 \pm 0.0053$ 	 	& $0.12 \pm 0.0086$ 		\\     
$ \theta_s$ & $1.04 \pm 0.0011$ 		  & 		$1.04 \pm 0.0033$ 		& $1.04 \pm 0.0011 $	\\ 
   $ n_s$ & $0.967 \pm 0.0095 $ 			& 		$0.979 \pm 0.013$ 		& $0.99 \pm 0.017 $		\\
   $\Omega_{e}$ &  $ < 0.014$ 		& 		$< 0.041 $			&  $< 0.013 $			\\  
   $ N_{eff}$ & -  					& - 							& $3.8 \pm 0.48$ 			 \\
   $ Y_{He}$ & $0.25 \pm 0.00015 $ 		& 		$0.25 \pm 0.00021$ 		& $0.26 \pm 0.0060 $	\\
   $ \Omega_{de}$ & $0.74 \pm 0.020$ 	& 		$0.71 \pm 0.029$ 		& $0.75 \pm 0.020$		\\
   $ H_0$ & $71.4 \pm 1.7$ 			& 		$69.5 \pm 2.3 $ 		& $76.0 \pm 3.5$ 		\\ 
\hline
\end{tabular}
\caption{Mean value (not best fit) plus standard deviation for a selection of baseline and derived parameters in EDE2. In the case of $\Omega_e$ and $w$ we report the $95\%$ confidence upper limit. Limits on $\Omega_e$ do not seem to be sensitive to marginalizing over the effective relativistic degrees of freedom $N_{eff}$.}\label{tab:ede2}
\end{table}  
\par\end{center}
\normalsize

\normalsize

\subsection{Parametrizations 3 and 4}
We now move to the runs using the parametrizations EDE3 and EDE4. For this test,
it is sufficient to fix $\tau$ in both cases, as we are only interested in
comparing results from EDE3 and EDE4 for different values of $a_e$ and $a_f$.
To do so, we consider EDE3 simulations for $a_e = (0.001, 0.01, 0.03, 0.1, 0.2)$
and EDE4 for $a_f = (0.003, 0.01, 0.03, 0.1, 0.2)$. 
In EDE3 we expect that for $a_e$ sufficiently small, we recover EDE2 results,
while for $a_e$ approaching $a_c$ we recover $\Lambda$CDM.
In EDE4 we expect that for $a_f$ small enough we recover $\Lambda$CDM while for
$a_f$ approaching $a_c$ we obtain EDE2.

\begin{figure*}
\centering
   \includegraphics[width=14.cm]{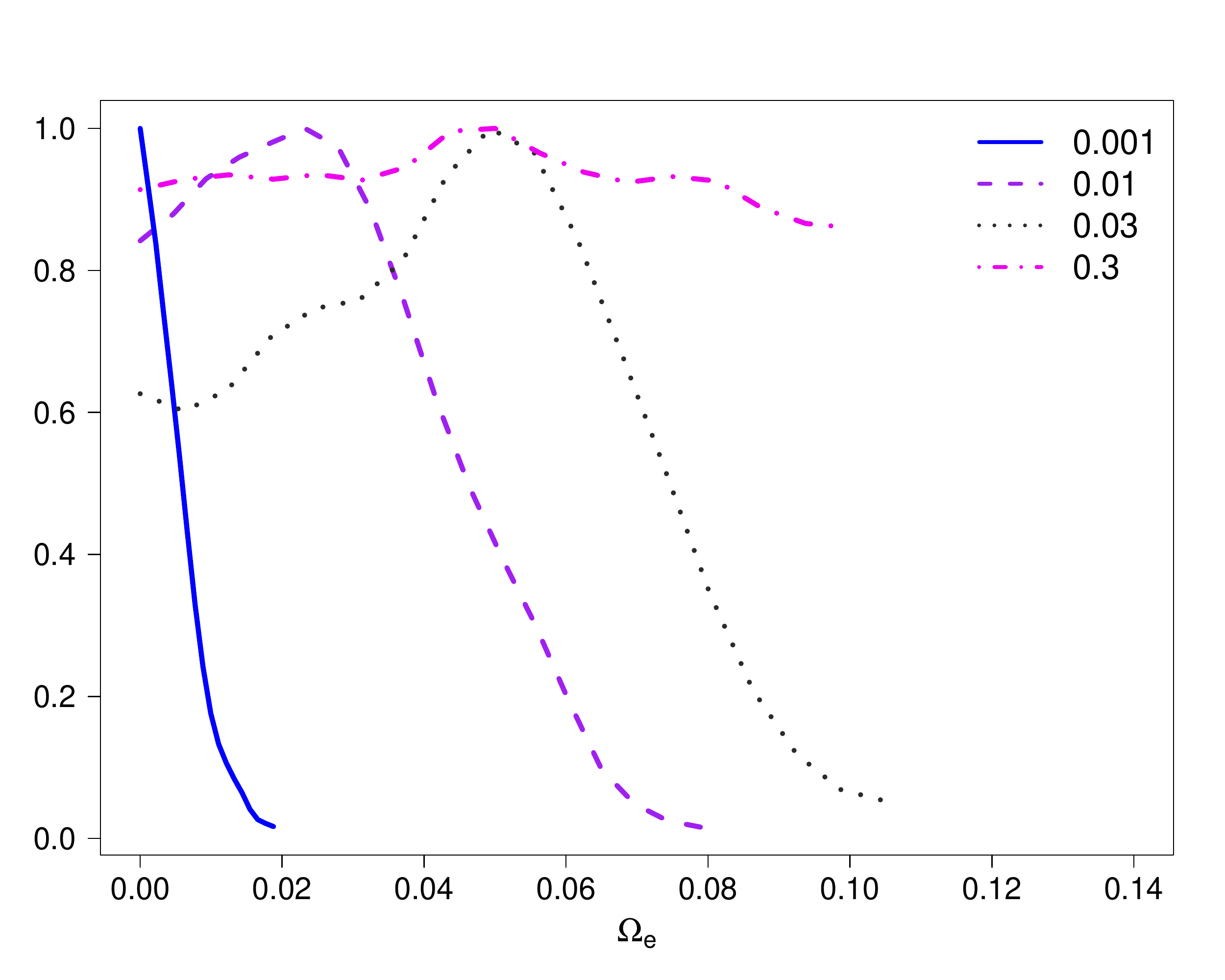}%
     \caption{\small 1D likelihoods for the cosmological parameters for EDE3 for
four values of $a_e$.
   As expected, for $a_e$ sufficiently large, we get no constraint on
$\Omega_{e}$ as we approach the $\Lambda$CDM scenario.}
\label{fig:like1D_par3}
\end{figure*}

In Fig.\ref{fig:like1D_par3} we show 1D likelihoods for different values of
$a_e$. For EDE3 we conclude that the bounds on $\Omega_e$ get weaker as the
onset of EDE is delayed. The presence or absence of dark energy in a redshift
range $z\approx 2-10$ is hard to be detected. 

\normalsize

\begin{figure*}
\centering
   \includegraphics[width=16.cm]{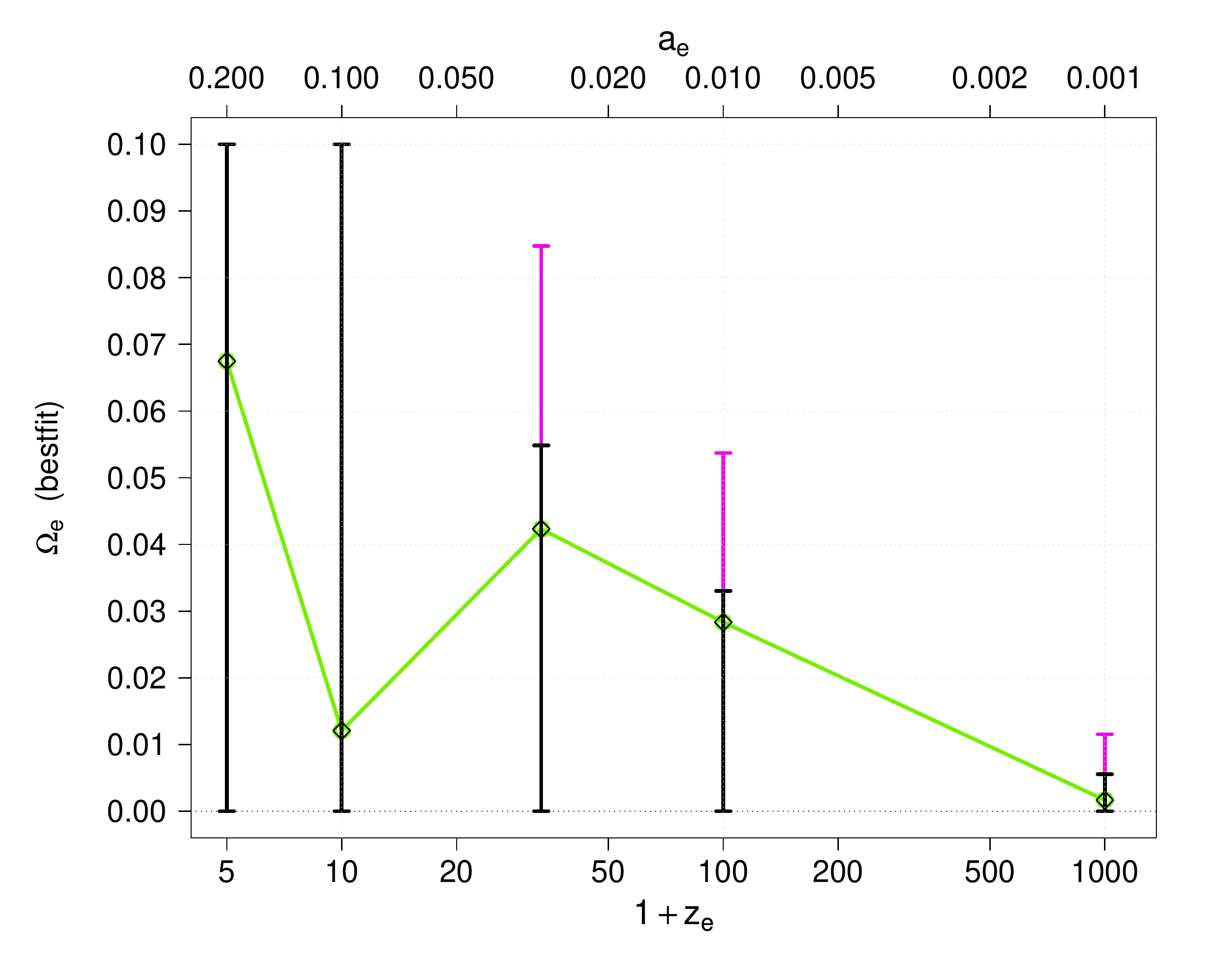}%
     \caption{\small Bestfit for $\Omega_e$ EDE3 as a function of $1 + z_e
\equiv 1/a_e$. Error bars correspond to 1 sigma (black) and 2 sigma (magenta)
marginalized limits (upper1 and lower1, upper2 and lower2) respectively. The
prior on $\Omega_e$ in the runs is $[0:0.1]$ so the error bars of the first two
points are saturated. This shows that for values of $z_e$ close to last
scattering, bounds are confirmed to be tight; for $1 + z_e \sim 5$, or
equivalently $a_e = 0.2$, we are close to a $\Lambda$CDM scenario; in the range
in between, for $a_e \sim 0.01-0.03$ (for which EDE becomes non negligible
around $z \sim 40-100$) $\Omega_e$ is less constrained and can be as high as
$8\%$ at 2 sigma from that time on.  The top axis shows the corresponding values
for $a_e$.}
\label{fig:par3_bestfits_z}
\end{figure*}

\begin{figure*}
\centering
   \includegraphics[width=12.cm]{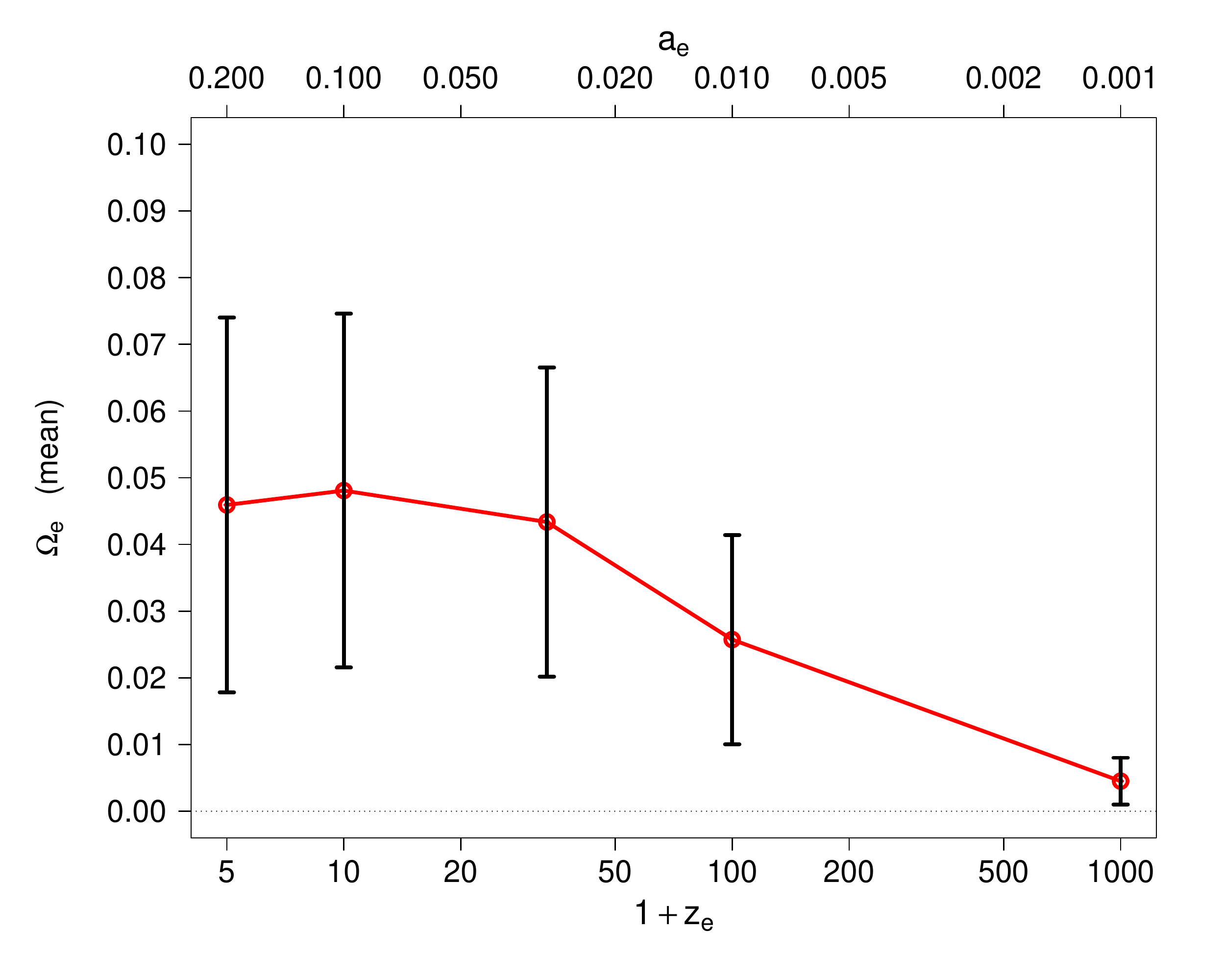}%
     \caption{\small Mean plus/minus standard deviation for $\Omega_e$ EDE3 as a
function of $1 + z_e \equiv 1/a_e$. }
\label{fig:par3_mean}
\end{figure*}

In Fig.\ref{fig:par3_bestfits_z} we show best fit results for EDE3, at
different values of $a_e$ or equivalently $1+z_e \equiv 1/a_e$. We recall that
$a_e$ indicates the time from which Dark Energy started to be non-negligible. We
see, as expected, that for values of $a_e$ close to CMB decoupling ($a \sim
0.001$) we approach limits found for EDE2. On the other side, for values of
$a_e$ closer to $a_c$, Dark Energy becomes non-negligible later, approaching a
$\Lambda$CDM scenario: in this case we finally have no limits on $\Omega_e$.
Interestingly, for the first time we show the limits we actually have at
intermediate $a_e$ or $z_e$: these become less stringent the later Dark Energy
starts to become non-negligible. In particular, for $a_e \sim 0.01-0.03$ (so if
EDE becomes non negligible around $z \sim 40-60$) $\Omega_e$ is allowed to be as
big as about $8\%$ at 2 sigma from that time on.
In Fig.\ref{fig:par3_mean} we also show the mean value of $\Omega_e$ for
different values of $a_e$. (For $z_e < 100$ the mean differs from zero at a bit
more than one sigma.)

\normalsize

\normalsize

\begin{figure*}
\centering
   \includegraphics[width=12.cm]{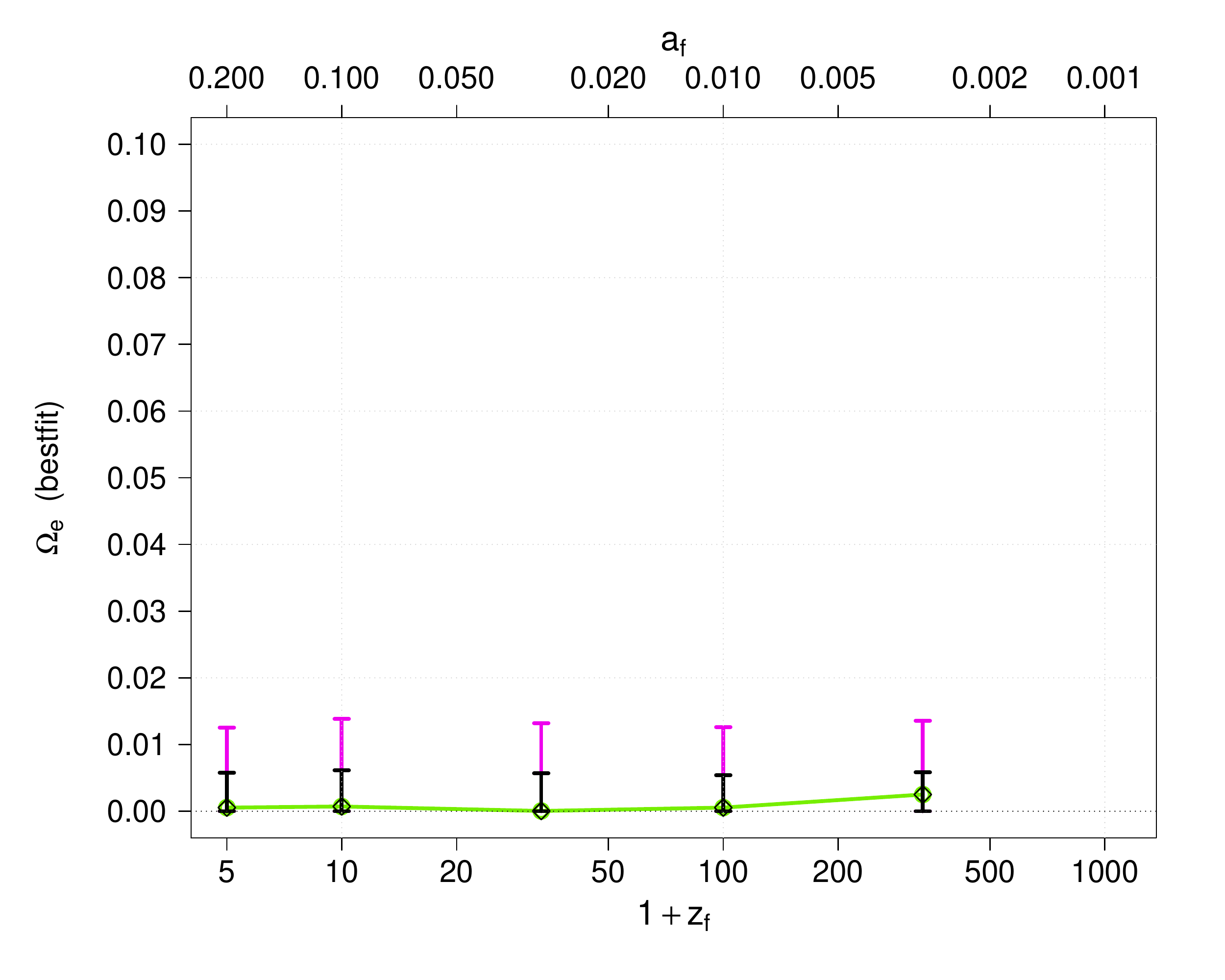}%
     \caption{\small  Bestfit for $\Omega_e$ EDE4 as a function of $1 + z_f
\equiv 1/a_f$. Error bars correspond to 1 sigma (black) and 2 sigma (magenta)
marginalized limits (upper1 and lower1, upper2 and lower2) respectively. This
shows that tight constraints on early dark energy come from CMB rather than
late time cosmology so that if a non negligible amount of dark energy is present
at CMB, constraints are almost independent of whether early dark energy
switches off afterwards. The main CMB constraints in presence of EDE do not come
from the suppression of the growth of structures or from the late ISW but rather
from its presence at last scattering. The top axis shows the corresponding
values for $a_f$.}
\label{fig:par4_bestfits_z_low}
\end{figure*}

  We recall that EDE3 and EDE4 are used here as a mean to isolate the two main
effects that are caused by the presence of EDE. The first is the reduced
structure growth in the period after last scattering; this implies a reduced
number of clusters as compared to $\Lambda$CDM and therefore a weaker effect of
the lensing potential on the CMB peaks. This first effect has been isolated
through parametrization EDE3, by switching on EDE3 only after $a_e$. The second
effect concerns the influence on the CMB peaks arising from the presence of dark
energy at the epoch of last scattering. This second effect is isolated in
parametrization EDE4, in which EDE only lasts until a final value $a = a_f$.
In Fig.{\ref{fig:par4_bestfits_z_low}} we plot the best fit results for EDE4, at
different values of $a_f$ or equivalently $1+z_f \equiv 1/a_f$. For EDE4 the
presence of dark energy during last scattering affects bounds at all times,
independently of $a_f$. This shows that tight constraints on early dark
energy come from the last scattering epoch rather than late time cosmology. Therefore 
 if a non negligible amount of dark energy is present at CMB, constraints
are almost  independent of whether early dark energy switches off afterwards.
The main CMB constraints in presence of EDE do not come then from the
suppression of the growth of structures or from the late ISW but rather from its
presence at last scattering.

\normalsize

\section{Conclusions} \label{conclusions}
A constant fraction of
early dark energy (EDE) reduces the growth rate over a large period and
therefore has a substantial effect on the power spectrum of cosmic structures or
the induced gravitational potential. If the growth rate is not affected by other
properties of a model one can place strong constraints on such EDE models. 
Furthermore, the presence of EDE at the time of last scattering modifies the geometry and therefore
influences the location and height of the CMB peaks. Again, this can be
constrained by precision CMB data at large multipoles.
For a
simple parametrization of the time evolution of dark energy involving only two
parameters, namely the fraction of dark energy at present
$\Omega_{de0}=1-\Omega_{m0}$, and the constant fraction of dark energy at early
times, $\Omega_e$, we find a constraint $\Omega_e<0.015$ at 95\% confidence
level. 

On the other hand, our detailed analysis reveals that the assumptions of
constant $\Omega_e$ and the absence of other factors influencing the growth rate
has a very strong impact on the constraint. If the onset of EDE is delayed,
starting only for a scale factor $a_e$, the bounds get much weaker,
$\Omega_e<0.08$ for $a_e\approx 0.02$. 
Furthermore, our analysis shows that the main constraints on EDE come from the last
scattering epoch rather than from the suppression of the subsequent growth of structure. We have
mimicked this by the EDE4 simulation for which dark energy is always present at
last scattering but becomes absent for $z < z_f$ that we have tested up to $1+
z_f = 300$ 
with no effect on the bounds. Our parametrization EDE4, compared to EDE1, shows
that even if the lensing potential is different for these two parametrizations,
constraints remain similar. This seems to indicate that gravitational lensing
is not the main source of information for Early Dark Energy bounds. This finding appears to be
in agreement with $case3$ of EDE1, in which lensing was switched off.

We  recall that realistic models in which early dark energy is present,
usually also have other effects which accompany the presence of a non negligible
amount of dark energy in the background. Typically, dark energy models also
affect clustering (ex. coupled quintessence, growing neutrinos, or any form of
'clustering' dark energy). The growth of structures will 
remain an important source of information constraining
 dark energy models via other probes such as weak lensing and baryonic
acoustic oscillations \cite{reviewdoc_euclid_2012, euclidredbook}.

Finally, we have proposed the new parametrization EDE2, similar to EDE1 but with
a sharper transition in $\Omega(a)$: this parametrization has the advantage of
having one parameter less than EDE1 ($w$ here is not a parameter). As bounds on
$\Omega_e$ do not depend substantially on how sharp the transition is, EDE2 is good enough to
picture a scenario in which early dark energy is present at all times. 
This may help to extract information about the time history of
dark energy from limited data sets. In view of Figs. \ref{fig:par3_bestfits_z} and \ref{fig:par3_mean}
it becomes clear, however, that bounds on Early Dark Energy depend crucially
on the precise time history or
 on \emph{how  early is early}.

\begin{acknowledgments}
VP is supported by Marie Curie IEF, Project DEMO. L.A. and C.W. acknowledge support from DFG through the TRR33 program
"The Dark Universe".
\end{acknowledgments}
\bibliography{pettorino_bibliography}

\end{document}